# IPOPv2 online service for the generation of opacity tables

Franck Delahaye
Carlo Maria Zwölf
Claude J. Zeippen

*LERMA, Observatoire de Paris, ENS, UPMC, UCP,PSL\*, CNRS, 5 Place Jules Janssen,
F-92195 Meudon Cedex, France*

Claudio Mendoza

*Centro de Física, Instituto Venezolano de Investigaciones Científicas (IVIC), PO Box
20632, Caracas 1020A, Venezuela*

**Abstract**

In the framework of the present phase – IPOPv2 – of the international Opacity Project (OP), a new web service has been implemented based on the latest release of the OP opacities. The user may construct online opacity tables to be conveniently included in stellar evolution codes in the format most commonly adopted by stellar physicists, namely the OPAL format. This facility encourages the use and comparison of both the OPAL and OP data sets in applications. The present service allows for the calculation of multi-element mixtures containing the 17 species (H, He, C, N, O, Ne, Na, Mg, Al, Si, S, Ar, Ca, Cr, Mn, Fe and Ni) considered by the OP, and underpins the latest release of OP opacities. This new service provides tables of Rosseland mean opacites using OP atomic data. We provide an alternative to the OPAL opacity services allowing direct comparison as well as study of the effect of uncertainties in stellar modeling due to mean opacities.

*Keywords:* Opacity, Atomic Data, Stars: Interior, Sun: Interior

## 1. Introduction

. The current IPOPv2[1] phase of the international Opacity Project (OP) and Iron Project aims at extending, improving and disseminating atomic data for physical and astrophysical applications. In this respect, the OP has produced radiative opacities for stellar studies which have been widely used in astrophysical applications. [See review by Zeippen [1] for an account of the first achievements of the OP and Badnell et al. [2] for a description of the present updated OP opacities including, in particular, inner-shell atomic transitions.]

. As part of the data services available at the Paris Astronomical Data Center (VOPDC[2]) as well as at the Centre de Données astronomiques de Strasbourg (CDS[3]), the atomic data for the 17 cosmic abundant elements (H, He, C, N, O, Ne, Na, Mg, Al, Si, S, Ar, Ca, Cr, Mn, Fe and Ni) used to calculate OP opacities can be retrieved from the TOPbase database [3]. Rosseland mean opacities and radiative accelerations can be computed from the online server referred to as the `OPserver` [4]. It must also be mentioned that these data services can be accessed through the VAMDC[4], the Virtual Atomic and Molecular Data Centre [5, 6]. In addition, a set of codes have been implemented which enable a sufficiently informed and careful user to exploit the data independently.

. The IPOPv2 group is concerned with the provision of several efficient and user-friendly channels for data dissemination in order to meet the increasing needs of the physical and astrophysical communities. The present report describes an efficient and versatile web service to deliver opacity data for a variety of modelling applications, which also facilitates comparisons between OP and other experimental or theoretical studies. The opacity table format we have adopted was first proposed by the OPAL team [7].

---

[1] http://cdsweb.u-strasbg.fr/topbase/
[2] http://vo.obspm.fr/services/
[3] http://cdsweb.u-strasbg.fr/OP.htx
[4] http://www.vamdc.eu/

. A word of caution is necessary at this stage. The OP work was planned to firstly produce accurate and complete atomic data sets, which would subsequently lead to the calculation of the monochromatic and mean opacities necessary in stellar envelope [8] and interior [2] studies, as well as the radiative accelerations required in asteroseismology [**?** ]. In the pursuit of this aim, compromises had to be made. Therefore, spectroscopic accuracy was not a primary objective for all transitions, and many weak transitions not essential for the computing of opacities were not included in the atomic database. Specific comparisons with data computed in other works for restricted temperature and density domains may thus reveal somewhat large discrepancies [9, 10], which do not invalidate the usefulness of the OP opacities for most astrophysical applications.

. Work by several teams, including the IPOPv2 group, is in progress to examine the possibilities of improving existing opacities. In this respect, detailed comparisons streamlined by practical data services will also be a powerful tool as laser physicists and astrophysicists require accurate opacities in their models.

. Recent studies of solar elemental abundances have questioned the accuracy of the currently available opacities. Asplund et al. [11] proposed an oxygen abundance which was notoriously lower than well-established values of Grevesse and Sauval [12], and downsized C and N abundances were also recommended. One of the ensuing consequences of this unexpected revision was to upset reliable helioseismological benchmarks, namely the depth of the convection zone, the helium surface content and the sound-speed profile. An easy way out of this difficulty was to advocate for increased stellar opacities. Even though the revisited abundances have increased again since 2004 [13, 14], the debate is still going on regarding the reliability of opacities and the integrity of the "new solar abundances" [15, 16, 17, 18, 19, 20].

. It is worth mentioning that helioseismological models have been performed with LEDCOP opacities [21] as well as with those from OPAL and OP showing general good agreement. However, the effects of the small standing discrepancies could be of more consequence when running an evolutionary code for which the easy access to multiple-source opacity tables could be of primary importance.

. Finally, experimental work is under way using powerful lasers in order to measure opacities. Some results and comparisons with theoretical data are already available [22, 10]. The OP opacities are not too far off from the new measurements when compared with other calculated results. While mean opacities for mixtures such as the solar case are certainly very accurate, line-by-line monochromatic opacities for certain elements may show differences with experiment [23, 24, 25, 26, 27]. More of this type of evaluations will become possible as experimentalists make progress in the near future.

## 2. Approximations

. The OP opacities uses only 17 elements in contrast to the OPAL tables; P, Cl, K and Ti are not included in the OP atomic data. Nevertheless, the log-scale form of the web service allows the inclusion of these elements in the mixture. Their corresponding fraction number is distributed to the nearest element present in OP data ($f_S = f_S + f_P$, $f_{Ar} = f_{Ar} + f_{Cl}$, $f_{Ca} = f_{Ca} + f_K$, $f_{Cr} = f_{Cr} + f_{Ti}$ ). This procedure as shown by Delahaye and Pinsonneault [15] produces negligible differences in the Rosseland mean, $\kappa_R$, as well as in the solar model properties (neutrinos fluxes, structures, etc.), and is better suited to uniformly distribute their fraction numbers to all elements.

. Ranges of physical conditions covered by OP are $\log(T) \in [3.5, 8.0]$ and $\log(\rho) \in [-15.0, 5.5]$. Such ranges correspond to $\log(R) \in [-8, 1]$, where $R = \rho/T_6^3$ with $T_6 = T/10^6$ is a useful parameter introduced by OPAL. However, the grid is not rectangular; for example, at $\log(T) = 3.5$ the density range spans $-15.0 \leq \log(\rho) \leq -4.5$ ($-12.5 \leq \log(R) \leq -2$ ) which is characteristic of stellar envelope conditions while, for the stellar deep-interior conditions of $\log(T) = 8.0$, we have $\log(\rho) \in [-2.0, 5.5]$ ($\log(R) \in [-4, 3.5]$).

. By convention in the OPserver, when data are required for $\rho-T$ points not covered by OP, an extremely large Rosseland mean opacity of $\log(\kappa_R) = 9.999$ is quoted. Generating tables with such large values at the missing points may cause bothersome differences in stellar modelling. As seen in Fig. 1, the interpolation routine included in the stellar modelling codes working with such tables generates a bump around $\log(T) = 5.5$. It is to be expected when the physical conditions reach the edge of the opacity table where the high value for $\kappa_R$ is set. In order to avoid such behavior, the 9.999 datum is replaced hereby with values extrapolated from the existing data.

. This is the sort of solution offered in the present OPtable data service, i.e. generating opacity tables without these extremely large values except at the very high temperatures $T$ where no data at all are available at any density. The extrapolation is made at a fixed $T$ using data from other density points. $\kappa_R$ is less sensitive to density changes than to temperature changes. Tests on the present linear extrapolation give errors under 1% except in some rare cases where they may reach 10%. When available in OPAL, the differences for such $\rho-T$ points for a typical solar composition (e.g. GS98) is of the same magnitude ($\sim 10\%$), while the general differences between OP and OPAL for the solar condition are less than 4% (cf. Delahaye and Pinsonneault [28]) although specific differences can reach 18% [2]. In fact differences are much larger for some $\rho-T$ points but they are inherent to the different approaches as shown in Badnell et al. [2].

. Results obtained from using the two types of online services are shown in Figs. 1–2. In these figures, we compare the extrapolated Rosseland mean opacity for a given stellar structure using two types of tables: one with the extrapolation and the other using the high $\kappa_R$ value (9.999). The reference is taken to be a fine-mesh table not yet included in the service. The extrapolation impact on the mean opacity may be clearly seen, specially the disappearance of the bump in Fig. 2 when compared to Fig. 1.

. We have noticed some differences (wiggles) at high temperatures depending on the resolution used in the opacity table. In OP, medium resolution corresponds to $\delta \log(T) = 0.05$. The fine-mesh grid ($\delta \log(T) = 0.025$) is required in order to remove the unwanted features at high $T$, especially for specific applications (e.g. the solar composition problem) in order to minimise the uncertainties in stellar modelling. Such tables will be provided later in the present service [29],and the routines to interpolate the opacity tables in stellar codes will require some minor changes in order to use them.

. A difference in the 6th digit of the relative abundances of one metal (elements other than H and He) generates a $\kappa_R$ variation under 0.25% in less than 5% of the points in the overall 126 tables provided in each OPtable file for a given composition. This accuracy level is better than the uncertainties due to the interpolation routines in stellar codes. As pinpointed by Bahcall et al. [30], such uncertainties are around 4%.

## 3. Service interface

### 3.1. The web GUI

. The OPtable home page can be accessed from the new IPOPv2 site[5], where the user may specify via a graphical user interface (GUI) elemental mixtures in one of two possible formats. The first type is similar to that in the OPAL page[6] where relative metal abundances are required. Inputs include the following elements: C, N, O, Ne, Na, Mg, Al, Si, S, Ar, Ca, Cr, Mn, Fe and Ni.

. The second format uses the logarithmic scale familiar to the stellar atmosphere community where $\log(\epsilon_{\mathrm{H}}) = 12$. It is based on 21 elements including P, Cl, K and Ti which are contained in the OPAL data sets. These elements are not taken into account in OP but, for the sake of simplicity, have been added to the input vector. However, the abundance for each of the four extra elements is transferred to the closest next element present in the OP database. The effects of such a procedure can be found in Delahaye et al. [18]. What will be henceforth referred to as $input_2$ consists then in providing $\log(\epsilon_{X_i})$ for $X_i$ in the vector {H, He, N, O, Ne, Na, Mg, Al, Si, P, S, Cl, Ar, K, Ca, Ti, Cr, Mn, Fe, Ni}. The mapping between these two input formats is controlled by

$$
\begin{cases}
\text{for } v_i \in input_1 \text{ and } w_j \in input_2 \\
v_i = \left[ 10^{w_i} \middle/ \left( \sum_{k=1}^{19} w_k \right) \right] \text{ for } i \in [1,8] \text{ or } i \in [13,15], \\
v_i = \left[ \left( 10^{w_i} + 10^{w_{i+7}} \right) \middle/ \left( \sum_{k=1}^{19} w_k \right) \right] \text{ for } i \in [9,12].
\end{cases}
\tag{1}
$$

---

[5]http://cdsweb.u-strasbg.fr/OP.htx
[6]http://adg.llnl.gov/Research/OPAL/type1inp.html

. A web page is dedicated to browsing, previewing and downloading existing tables. Every table computed with the service is stored (in an anonymous way, i.e. with no user reference) to avoid unnecessary recalculations. Tables are labeled with the scheme

$OP17.$[C]-[N]-[O]-[Ne]-[Na]-[Mg]-[Al]-[Si]-[S]-[Ar]-[Ca]-[Cr]-[Mn]-[Fe]-[Ni].$stored$

where $[X]$ is the $X$-element fraction used for processing a given table with the $input_1$ format. A second web page provides instructions for downloading, installing and using the stand-alone client.

### 3.2. The JSP layer

. The *Java Servlet Page* (JSP) layer processes the information posted by users to the server via the GUI. Starting from the latter, it generates a set of scripts and parameter files for handling and executing the Fortran procedures. It is in this layer that, according to user needs and choices, the input mapping described by equation (1) is performed.

. The computation of a given opacity table is highly parallelized: the acceleration is linear up to 126 processors[7]. Moreover, it is important to notice that, as a result of their design and architecture, the scripts from different users are processed simultaneously with no concurrency problem regarding inputs and outputs.

### 3.3. The computation layer

The computation layer executes the scripts generated in the JSP layer. As previously mentioned, scripts from different users can be processed simultaneously. However, since in our server configuration the computation of a single table uses the CPU resources completely, we have implemented a scheduler based on the FIFO[8] principle: all the jobs are queued and singularly executed in arrival order.

Once the processing is completed, results are sent by email to the user and stored in the archive area, where computed tables can be browsed via the GUI.

### 3.4. The stand-alone client

By means of the web interface, a user must manually fill a web form for each table request. This is not user-friendly, especially for processing many requests. With this problem in mind, we have written a stand-alone client that enables the handling of multiple mixtures simultaneously.

The client is a Java graphical application that users can easily run on their desktops. Each mixture to be processed must be previously stored in a dedicated text file (cf. Table 1 for details of the format of such composition files). As shown in Table 1, the client can handle both $input_1$ and $input_2$ formats without user assistance. By running the client, users can choose a set of comp files (in both formats) locally. For each file in the set, the client will automatically invoke the service (i.e. the JSP layer of the remote server), and the sequence of steps follows the standard server workflow. Users will receive results by e-mail (a mail for each mixture). Moreover, no computation is performed on the client side: the client is just an alternative rich user interface to the web page.

---

[7]The server currently hosting the service is a 7-core Intel Xeon 2.5 Ghz. One core is reserved for the web server and six are used for computations. Table processing on this 6-core processor takes approximatively 4 minutes

[8]First in, first out

### 3.5. Further developments

The development of the present facility has been an opportunity to introduce some new ideas for defining constraints and performing parameter verifications in scientific web services. These ideas have led to the definition of a new standard in the International Virtual Observatory Alliance: the *Parameter Description Language* [31]. With the PDL grammar, one can precisely describe all the service inputs and outputs (with their related constraints) into a machine actionable way. Each PDL description takes the form of a XML file. A condition such as

$$\left\{ \begin{array}{l} \text{If } f(p_1) > g(p_2) \\ (p_i \ f \text{ and } g \text{ being parameters and functions}) \\ \text{then } p_3 \text{ must satisfy a given mathematical condition.} \end{array} \right. \tag{2}$$

can be easily expressed using the PDL and univocally serialised into an XML file. Since all the parameters (and related constraints) are described in a machine actionable form, the PDL framework automatically generates from the description

- a dynamic graphical client,

- the checking algorithms for validating user provided values,

- the interoperability graphs between all the services described using PDL.

We have therefore successfully provided the IPOPv2 service with a PDL overlayer. The description[9] is based on the first input format (cf. Section 3.1) and

---

[9]The XML serialisation of this description is available at the url: http://pdl.obspm.fr/descriptions/opacity.xml

specifies the basic constraints

$$\sum_{X \in El} [X] = 1 \tag{3}$$

where

$$El = \{C, N, O, Ne, Na, Mg, Al, Si, S, Ar, Ca, \quad Cr, Mn, Fe, Ni\} \ . \tag{4}$$

By processing this description, the PDL framework generates a new ad hoc client (alternatively to the first described in Section 3.4). This client embeds the automatically generated code verifying the above sum condition.

The IPOPv2 service has been successfully used for testing and validating the PDL framework. The PDL layer could be used in the future for interconnecting, in a fully interoperable way, this service with existing atomic data bases and other computational services.

## 4. Conclusions and future work

We describe a new user-friendly service which provides an efficient and practical method to access detailed opacity tables. The service will be improved in the future on the basis of progress made by data producers, data base experts and, of course, the reactions and needs of data users.

We provide a new service which has been thoroughly tested since 2012. It allows for the generation of opacity tables directly usable in unmodified standard stellar codes. The tables have the same structure and format as the OPAL tables. There are no constraints on the element compositions and one can prepare any table using his own mixture.

One next step in the development of IPOP services will be the extension of the table to a fine grid ($\Delta logT = 0.025$ and $\Delta \ log \ \rho = 0.25$) Orban et al. [29]. This will reduce the uncertainties introduced by the interpolation routines present in the stellar codes.

Another development we are working on is the production of equation of state (eos) tables. We will provide eos tables consistent with the opacity tables.

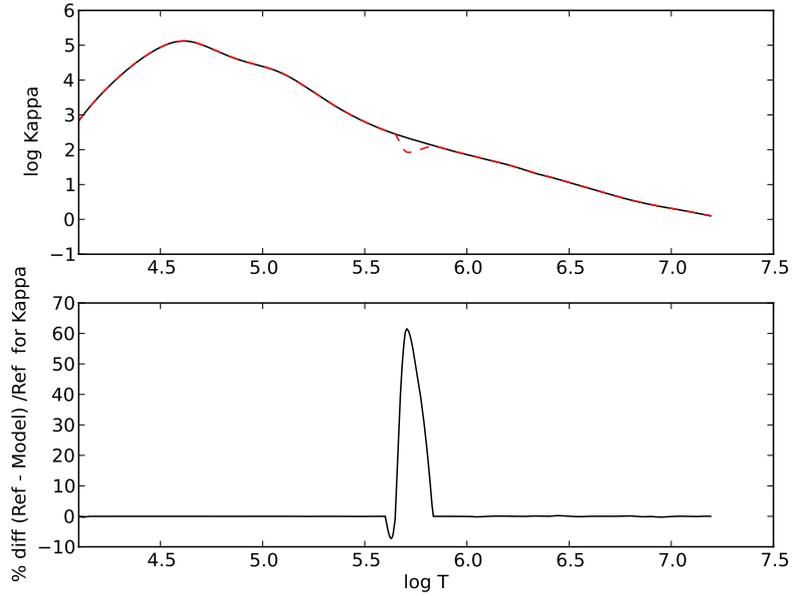

Figure 1: Log of Rosseland mean $log(\kappa_R)$ (top panel) and differences in $\kappa_R$ (bottom panel) in the sens of (extended table - original table) for the solar structure at zams. The differences arise from the presence of 9.999 values in the original opacity tables when no extrapolation is performed for these points.

The OPAL format will be adopted in order to avoid any change in stellar codes. The available options for the compositions will be of the same kind as the present ones and interpolation between tables will not be necessary since direct calculations for any mixture will be possible [32].

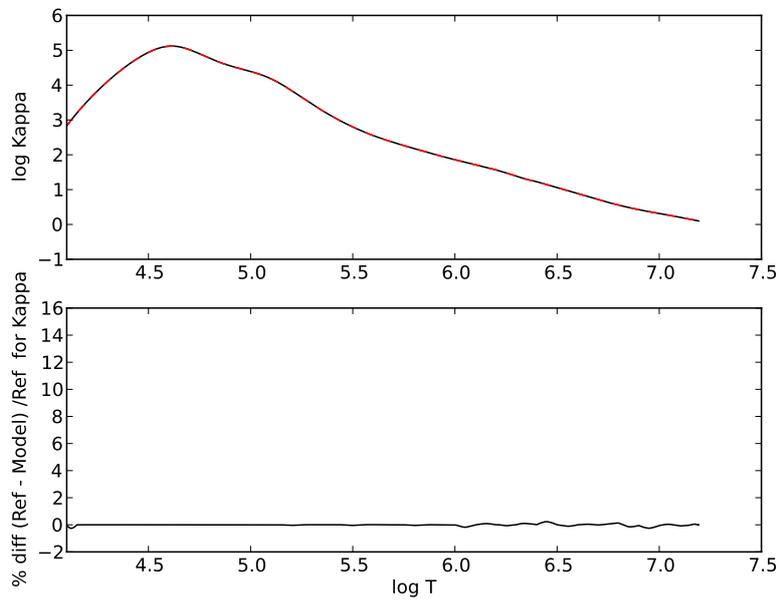

Figure 2: Log of Rosseland mean $log(\kappa_R)$ (top panel) and differences in $\kappa_R$ (bottom panel) in the sens of (extended table - original table). The extrapolation (see Section 2 for details) removes the spurious values for $\log(T) \in [5.5, 6.0]$. The wiggles at high $T$ are due to the $T$ resolution. We have used a medium-size mesh, but these features are removed from tables using a fine mesh.

Figure 3: Type-1 form for specifying the metal content of the requested mixture by means of fraction numbers.

Figure 4: Type-2 form for specifying the metal content of the requested mixture on a log scale.

Table 1: Details of files structures handled by standalone client. Note: For the $input_2$ format the extra element (P, Cl, K and Ti) are not present in the OP data. Their abundance are transfered to the next closer element. (See text for details).

| Line number | $input_1$ format | $input_2$ format |
|:---:|:---:|:---:|
| 1 | [C] | $log\ \epsilon_C$ |
| 2 | [N] | $log\ \epsilon_N$ |
| 3 | [O] | $log\ \epsilon_O$ |
| 4 | [Ne] | $log\ \epsilon_{Ne}$ |
| 5 | [Na] | $log\ \epsilon_{Na}$ |
| 6 | [Mg] | $log\ \epsilon_{Mg}$ |
| 7 | [Al] | $log\ \epsilon_{Al}$ |
| 8 | [Si] | $log\ \epsilon_{Si}$ |
| 9 | [S] | $log\ \epsilon_S$ |
| 10 | [Ar] | $log\ \epsilon_{Ar}$ |
| 11 | [Ca] | $log\ \epsilon_{Ca}$ |
| 12 | [Cr] | $log\ \epsilon_{Cr}$ |
| 13 | [Mn] | $log\ \epsilon_{Mn}$ |
| 14 | [Fe] | $log\ \epsilon_{Fe}$ |
| 15 | [Ni] | $log\ \epsilon_{Ni}$ |
| 16 | - | $log\ \epsilon_P$ |
| 17 | - | $log\ \epsilon_{Cl}$ |
| 18 | - | $log\ \epsilon_K$ |
| 19 | - | $log\ \epsilon_{Ti}$ |